\documentclass[conference]{IEEEtran}
\IEEEoverridecommandlockouts

\usepackage{cite}
\usepackage{amsmath,amssymb,amsfonts}
\usepackage{algorithmic}
\usepackage{graphicx}
\usepackage{textcomp}
\usepackage{xcolor}
\usepackage{stfloats}
\usepackage{float}
\usepackage{tabularx}
\usepackage{multirow}
\usepackage{caption}
\usepackage{subfig}
\usepackage{setspace}
\usepackage{graphicx}
\usepackage{amsmath}
\usepackage{amsfonts}
\usepackage{amssymb}
\usepackage{url}
\usepackage{hyperref}
\usepackage{xcolor}
\usepackage{colortbl}
\usepackage{booktabs}
\usepackage{placeins}

\usepackage{titlesec}
\titlespacing*{\section}{0pt}{3.0ex plus 1ex minus .2ex}{2.0ex plus .2ex}
\titlespacing*{\subsection}{0pt}{2.0ex plus 1ex minus .2ex}{2.0ex plus .2ex}

\graphicspath{{Figures/}}

\usepackage{amsthm}  

\theoremstyle{plain}

\theoremstyle{definition}

\theoremstyle{remark}

\theoremstyle{definition}

\def\BibTeX{{\rm B\kern-.05em{\sc i\kern-.025em b}\kern-.08em
    T\kern-.1667em\lower.7ex\hbox{E}\kern-.125emX}}
\begin{document}

\title{Can Physician Expertise Improve Machine Learning Identification of Delirium?\
}

\author{%
\IEEEauthorblockN{Xinyu Qin$^{\dagger}$, Vicky Ye$^{\ddagger}$, Ruiheng Yu$^{\dagger}$, and Lu Wang$^{\dagger,\dagger\dagger,*}$}

\IEEEauthorblockA{%
$^{\dagger}$\textit{Department of Biomedical Engineering, University of Houston}, USA\\
\texttt{\{xqin5, ryu11\}@cougarnet.uh.edu, lwang71@central.uh.edu}\\
$^{\dagger\dagger}$\textit{Department of Health Systems \& Population Health Sciences, University of Houston}, USA\\
$^{\ddagger}$\textit{Paul G. Allen School of Computer Science \& Engineering, University of Washington}, USA\\
\texttt{vickyye@cs.washington.edu}%
}

\thanks{$^{*}$Corresponding author: Lu Wang (lwang71@central.uh.edu).}%
}

\maketitle

\begin{abstract}
Delirium is common in hospitalized patients and is often missed in routine care. We present a user-centered interactive machine learning (UC-iML) framework for delirium detection support that combines physician-guided feature refinement with interpretable modeling. Using 3,862 labeled admissions from six Toronto hospitals in the General Medicine Inpatient Initiative (GEMINI), we integrate administrative variables, laboratory results, medications, and a radiology-derived text indicator. Physicians guide feature refinement and model evaluation, and Shapley Additive exPlanations (SHAP) are used to summarize feature attribution. We evaluate standard supervised classifiers with temporally separated holdout testing and a later-phase validation cohort. Compared with automated and baseline variants, the proposed framework shows better overall discrimination and stronger temporal robustness, while the explanations highlight clinically meaningful signals. These results support UC-iML as a practical human-in-the-loop framework for clinically relevant delirium modeling.
\end{abstract}

\section{BACKGROUND}
Delirium, termed “acute brain failure,” is a neuropsychiatric condition labeled as both a “medical emergency” and a “silent epidemic,” affecting up to 50\% of elderly hospitalized patients, with symptoms like confusion, agitation, and hallucinations causing significant distress to patients and caregivers \cite{maldonado2017, bruera2009}. Patients with delirium face over twice the risk of in-hospital death or nursing home placement, heightened cognitive decline, dementia risks, longer hospital stays, increased readmissions, and doubled healthcare costs, with U.S. expenditures reaching \$183 billion annually \cite{leslie2008,hshieh2018}.


Up to 40\% of delirium cases are preventable through multicomponent programs, which save approximately \$16,000 USD per person-year post-episode \cite{inouye1999}. However, these programs are underused due to the lack of standardized protocols \cite{teodorczuk2012}. Scalable early detection tools, such as explainable machine learning (ML) models, are crucial for improving care and using delirium as a quality indicator.

Existing prediction models often rely on non-routinely collected data, such as cognitive and sensory impairment, which limits scalability \cite{mccoy2016underreporting}. For instance, the UK's National Institute for Clinical Excellence (NICE) model requires said data to be available in electronic records \cite{rudolph2016validation}. They also suffer from low accuracy, limited ML methods, neglect of text data, and issues like model drift \cite{oh2018prediction,corradi2018prediction}. Explainable artificial intelligence (XAI) methods like Shapley Additive exPlanations (SHAP) help clearly identify predictive features and monitor performance shifts \cite{lindroth2018systematic}. 

This paper analyzes the General Medicine Inpatient Initiative (GEMINI) dataset, containing data from over 370,000 patient admissions in 30+ Ontario hospitals. A subset of nearly 4,000 labeled admissions from six Greater Toronto Area hospitals includes 25\% delirium cases. The study spans two phases: phase 1 (April 2010-March 2015) and phase 2 (January 2018-January 2020). 

Key contributions are:
\begin{itemize}
    \item A user-centered interactive machine learning (UC-iML) workflow that incorporates physician input into feature refinement, evaluation, and interpretation.
    \item A temporal and cross-phase evaluation on GEMINI showing practical robustness under dataset shift.
\end{itemize}

\section{THE ROLE OF INTERACTIVE MACHINE LEARNING (iML)}
\vspace{-0.5pt}
While artificial intelligence (AI) has revolutionized radiology and pathology, its application in clinical care remains limited due to insufficient digitized data and computational methods \cite{raghupathi2014, hinton2018deep}. Additionally, much healthcare AI development relies on a few large, de-identified public datasets like MIMIC \cite{saeed2002mimic}. Automated ML (aML) methods, though common in image and speech tasks, are resource-intensive and lack transparency due to black-box approaches. Interactive ML (iML) addresses these issues by involving domain experts throughout the development process to enhance feature selection, interpretability, and adoption \cite{holzinger2019interactive, luo2025hai, li2026prefix, luo2026agentauditor}.

\begin{figure}[t]
    \centering
    \subfloat[A typical iML process for a medical application with a Human--Computer Interaction (HCI) interface.]{%
        \includegraphics[width=\columnwidth]{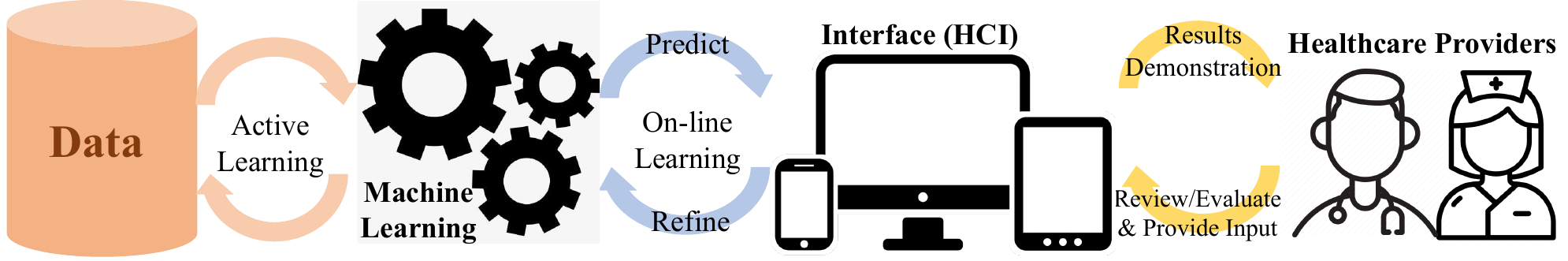}
        \label{appx:iml}
    }
    \vspace{3mm}

    \subfloat[A generalized framework of UC-iML consisting of three iterative stages: Requirements Acquisition (RA), Design \& Prototyping (DP), and Machine Teachers-Centered Evaluation (MTCE).]{%
        \includegraphics[width=\columnwidth]{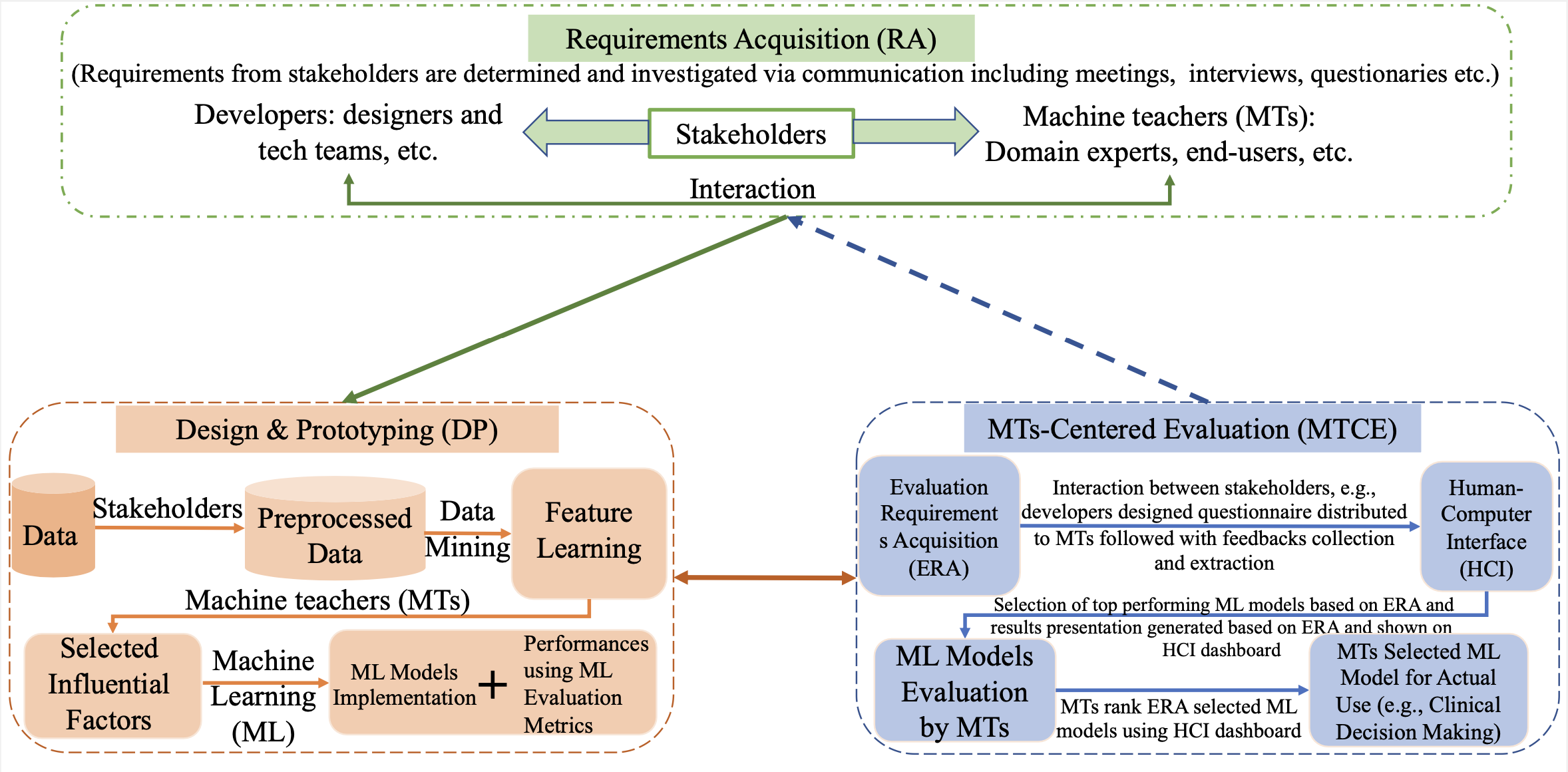}
        \label{guciml}
    }

    \caption{iML frameworks for healthcare applications. (a) illustrates a typical iML workflow from data to healthcare providers, while (b) presents the proposed UC-iML framework with three iterative stages integrating physician expertise and stakeholder interaction throughout model development.}
    \label{fig:combined_iml}
\end{figure}

\label{typical-iML-process} In typical healthcare iML processes, healthcare professionals are conventionally only involved after data preparation, ML model development, and prediction (see Fig. \ref{fig:combined_iml}a). Extending their active involvement throughout the process improves model performance, trust, and clinical adoption among decision-makers and stakeholders.

Inspired by human-in-the-loop designs, this study proposes a UC-iML framework consisting of three iterative, well-defined stages, as illustrated in Fig. \ref{fig:combined_iml}b: 

\begin{enumerate}
    \item \textbf{Requirements Acquisition (RA)}: Stakeholders define requirements, with machine teachers (MTs) contributing domain expertise through meetings and forms. 
    \item \textbf{Design and Prototyping (DP)}: Developers create prototypes, iteratively refining them with MT feedback. 
    \item \textbf{Machine-Teachers-Centered Evaluation (MTCE)}: MTs evaluate model performance using dashboards, selecting the final model for clinical use.
\end{enumerate}

This framework aims to integrate physician expertise into ML to improve delirium prediction and establish a replicable pipeline for clinical applications.
\vspace{-0.5pt}

\section{THE GEMINI-DELIRIUM DATASET}
\vspace{-0.5pt}
\subsection{GEMINI Study}
\vspace{-0.5pt}
GEMINI is a multi-institutional research collaboration in Ontario, Canada, standardizing electronic clinical data from six major hospitals (St. Michael’s Hospital, Toronto General Hospital, Toronto Western Hospital, Trillium Credit Valley Hospital, Trillium Mississauga Hospital, and Sunnybrook Hospital). Operational since 2017, it initially covered 240,000 hospitalizations to general internal medicine from 2010-2017. GEMINI now encompasses over 500 million data points on 400,000 hospitalizations, supporting AI-driven analytics with validated accuracy of 98-100\% across key data types \cite{verma2021assessing}.
\vspace{-0.5pt}
\subsection{Research Ethics Review Board (REB) Approval}
\vspace{-0.5pt}
The Toronto Academic Health Science Network's REB approved on the GEMINI study on 08/31/2019 (reference 15-087), with an extension granted on 8/17/2020. This paper is also part of a GEMINI sub-study on AI-based delirium prediction, approved by the University of Toronto REB on 10/15/2019 (RIS 38377) and renewed on 09/01/2020 under the same reference.
\vspace{-0.5pt}
\subsection{GEMINI Dataset}
\vspace{-0.5pt}
The GEMINI dataset links administrative and clinical data at the patient level:
\begin{itemize}
    \item Administrative Data: Comprehensive patient characteristics from the Canadian Institute for Health Information Discharge Abstract Database (CIHI-DAD) and National Ambulatory Care Reporting System (NACRS). Diagnoses and interventions systematically coded using Canadian International Statistical Classification of Diseases and Related Health Problems (ICD-10-CA) and the Canadian Classification of Health Interventions.
    \item Clinical Data: Includes lab results, transfusions, medications, vital signs, room transfers, and clinical monitoring. Statistical quality controls ensure data reliability \cite{verma2021assessing}. Radiologist reports enable text-based analytics.
\end{itemize}

\begin{figure}[t]
    \centering
    \includegraphics[width=0.95\columnwidth]{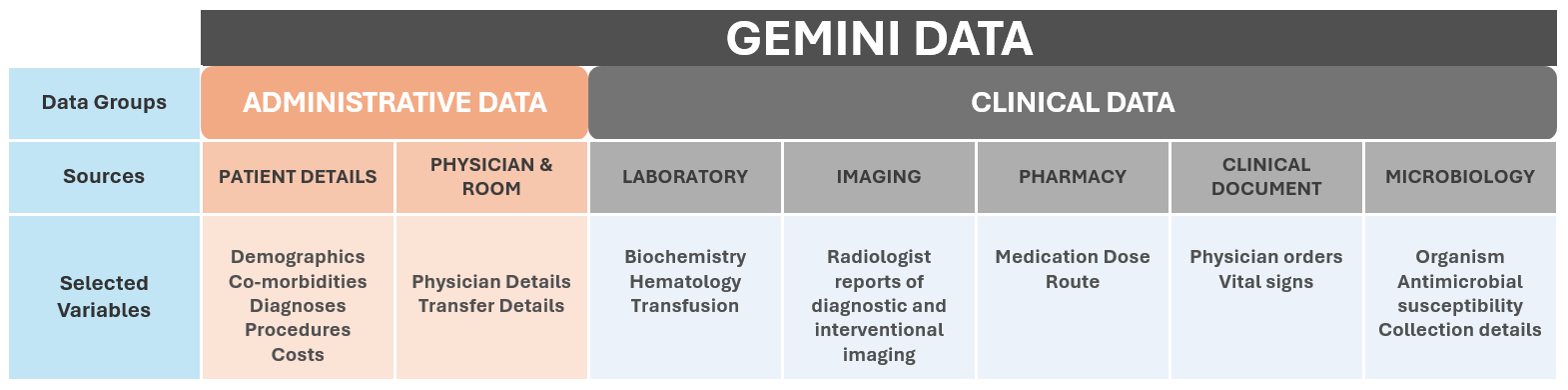}
    \caption{Data contained in a GEMINI project.}
    \label{data_in_GEMINI}
\end{figure}

Delirium cases are identified using a validated method with 75\% sensitivity and 83\% specificity, achieving 90\% inter-rater reliability for 5\% of independently reviewed charts \cite{raghupathi2014, inouye2005chart}. This study uses 3,862 labeled hospital admissions, combining clinical and administrative data from eleven GEMINI files. See Fig.~\ref{data_in_GEMINI} for detailed data descriptions.

\vspace{-0.5pt}
\section{USER-CENTERED iML FRAMEWORK APPLIED IN GEMINI-DELIRIUM}
\vspace{-0.5pt}

The UC-iML framework supports developing delirium prediction models via three iterative stages: RA, DP, and MTCE, with repeated clinician--developer feedback loops throughout the process (Fig.~\ref{guciml}).

Domain experts use clinically meaningful groupings (e.g., ICD-10 categories~\cite{verma2018prevalence}) to guide
feature prioritization and reduce dimensionality. In MTCE, physicians review key model trade-offs (screening utility, calibration, and temporal stability) and provide feedback that informs subsequent iterations, without relying on any single metric for selection.




\vspace{-0.5pt}
\subsection{Requirements Acquisition (RA)}
\vspace{-0.5pt}

In the RA stage, ML developers collaborated with physicians (MTs and end-users) to define delirium properties in GEMINI and translate clinical priorities into model requirements. Physicians prioritized influential factors spanning both administrative and clinical sources, including patient characteristics (age, gender, length of stay, discharge disposition, death, intensive care unit (ICU) admission, Charlson score), care-process signals (number of room transfers), laboratory indicators (occurrence of uric acid, vitamin B12, and thyroid-stimulating hormone (TSH) tests; abnormal sodium, creatinine, white blood cell (WBC), and troponin; albumin; Laboratory-based Acute Physiology Score (LAPS score)), and medication patterns (antibiotics, antipsychotics, benzodiazepines, cognitive enhancers, sedative-hypnotics, and total medication orders).

Early 5-fold cross-validation shows limited clinical utility (accuracy=0.74, sensitivity=0.51, precision=0.65, F1 score=0.56, area under the receiver operating characteristic curve (ROC-AUC)=0.77), and physicians noted that accuracy alone can be misleading because a trivial no-delirium predictor achieves a comparable baseline (about 75\%). The agreed requirements are to prioritize high sensitivity for screening, flag high-certainty cases when complete records are available, model delirium onset using admission-to-onset data, and report prospective validation with basic bias checks.

\vspace{-0.5pt}
\subsection{Design and Prototyping (DP)}
\vspace{-0.5pt}

During DP, physicians guided data inclusion and clinically meaningful feature construction, while the ML team implemented the pipeline. The contribution here is not a new classifier, but a user-centered workflow that combines multi-source feature engineering, physician-guided refinement, and standardized temporal evaluation.

\textbf{Comparators.} We compare the proposed method against two ablations: (i) \textit{aML-Delirium}, which applied data-driven feature selection without physician input, and (ii) \textit{Baseline-Delirium}, which excluded text-derived features and physician-guided refinement. All three settings used the same 3,862 admissions, differing only in feature construction and selection (feature counts: 324, 301, and 538, respectively).


\textbf{Feature construction and multi-source integration.}
We construct admission-level features from laboratory tests (summary statistics and test-performed indicators), diagnosis groups (ICD-10 mapped to Clinical Classification Software (CCS) categories) \cite{verma2018prevalence}, procedures and transfers (counts), clinical risk scores (Charlson, LAPS, Kidney Disease: Improving Global Outcomes (KDIGO)) \cite{quan2011updating,escobar2008riskadjusting,khwaja2012kdigo}, emergency department (ED) triage score \cite{bullard2014revisions}, medications, and administrative variables (service type, discharge destination, and length of stay). Radiology reports (CT/MRI) are concatenated per admission and summarized into a single text-derived indicator using term frequency–inverse document frequency (TF-IDF) features and a tuned gradient boosting classifier \cite{devika2016sentiment,scikit-learn}. 

\textbf{Expert knowledge encoding.}
After lasso-based screening, physicians refined feature retention to preserve clinically meaningful predictors and etiologically relevant signals, particularly laboratory abnormalities and radiology-derived indicators used in delirium assessment \cite{american1999practice,trzepacz2004neuropsychiatric}.

\begin{figure}[t]
    \centering
    \includegraphics[width=0.95\columnwidth]{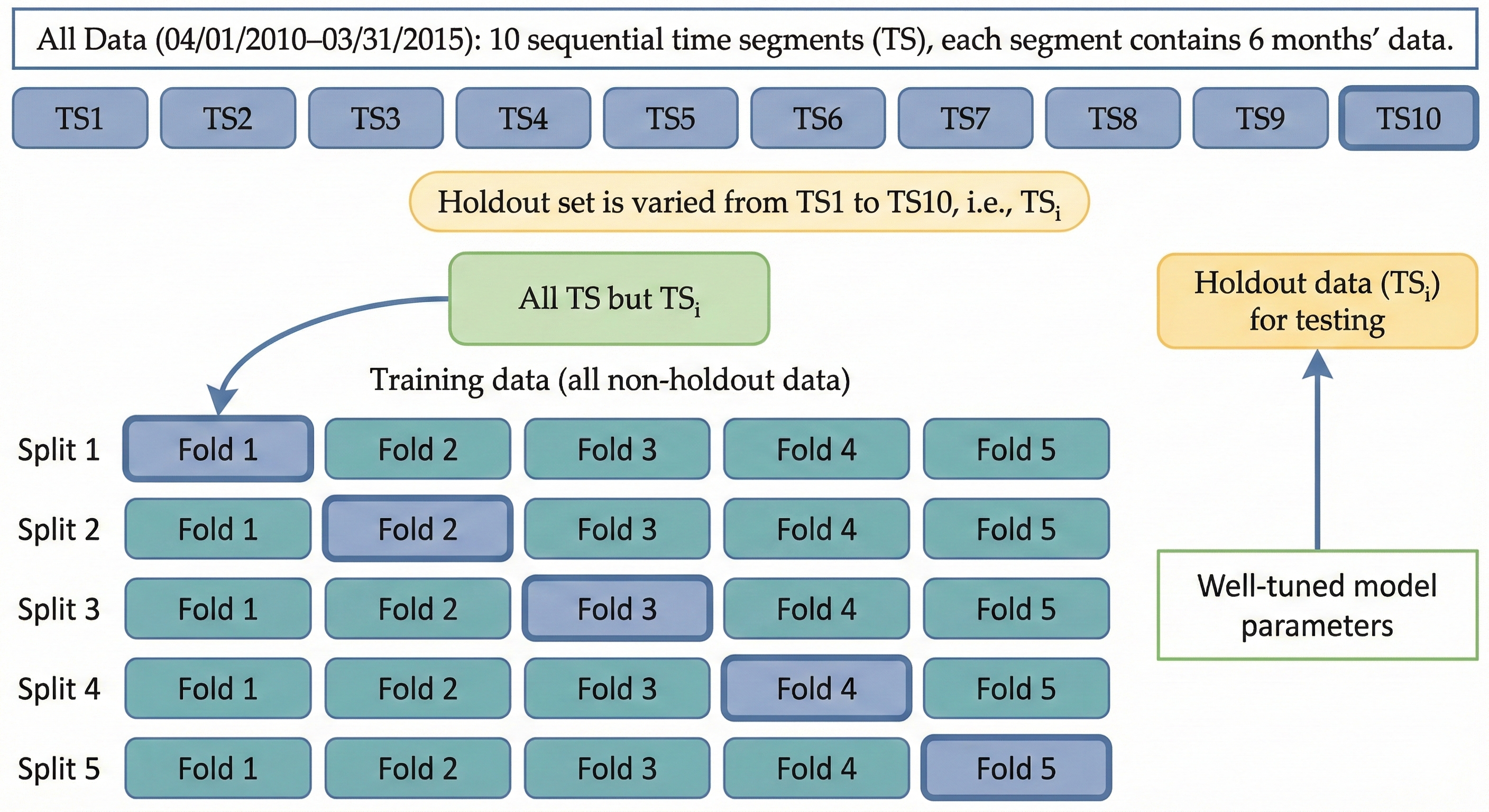}
    \caption{Rolling holdout evaluation pipeline across 10 6-month time segments.}
    \label{tt2}
\end{figure}

Phase~1 data (Apr 2010--Mar 2015) are segmented into 10 consecutive 6-month time segments (TS1–TS10), as shown in Fig.~\ref{tt2}, where each TS is alternately held out for testing while the remaining segments are used for training with internal cross-validation. Models are trained on the first 9 intervals and evaluated on the final interval (Oct 2014--Mar 2015) as a prospective holdout set. In addition, rolling holdout evaluation across the ten intervals assessed temporal robustness.

\textbf{Cross-phase drift test.}
To assess dataset shift, we evaluate the Phase~1-trained model on Phase~2 data from one hospital with 358 labeled admissions (Jan 2018--Jan 2020), reporting averaged 5-fold cross-validation results.

From the 12 classifiers, the top-performing models are Gradient Boosting Classifier (GBC), AdaBoost Classifier (ABC) and Logistic Regression (LR); GBC achieved the best F1 score and is selected.

\textbf{Interpretability.}
We use SHAP for feature attribution. For feature $j$, the Shapley value $\phi_j$ is
{\small\begin{equation}
\phi_j \;=\; \sum_{S \subseteq \{1,\ldots,p\}\setminus\{j\}} 
\frac{|S|!\,(p-|S|-1)!}{p!}\,
\Bigl[\,\mathrm{val}(S \cup \{j\}) - \mathrm{val}(S)\,\Bigr],
\end{equation}}
where $p$ is the number of features, $S$ denotes a subset of features, and $\mathrm{val}(\cdot)$ is the model value function. SHAP values quantify each feature's contribution to the prediction relative to the dataset baseline, supporting clinical interpretation of dominant risk factors.

\vspace{-0.5pt}
\subsection{Machine Teachers-Centered Evaluation (MTCE)}
\vspace{-0.5pt}
To ensure clinical validity and support model selection, physicians 
evaluated candidate models through an interactive dashboard developed 
in our previous work~\cite{Wang2022Physician}. The evaluation framework 
assessed three key dimensions: (1) discrimination performance across 
eight standard metrics (accuracy, precision, recall, F1-score, specificity, 
ROC-AUC, false alarm rate, miss rate); (2) temporal stability quantified 
via standard deviation across the 10 time segments; (3) calibration 
quality through reliability diagrams comparing predicted probabilities 
with observed outcomes.

Physicians prioritized high recall to minimize missed delirium cases, 
followed by balanced F1-score, reliable calibration, and temporal stability. 
Based on these criteria, GBC is 
selected for clinical deployment. GBC achieved holdout test performance 
of F1=0.719 (accuracy=0.851, ROC-AUC=0.902) with stable temporal behavior 
(F1 SD=0.024 across segments, range=0.646--0.719) and well-calibrated 
probability estimates. The tree-based ensemble structure enabled interpretable feature importance inspection for clinical review while maintaining superior predictive performance.
\vspace{-0.5pt}
\subsection{Commonly Used ML Model Evaluation Metrics}
\vspace{-0.5pt}
Delirium model performance is assessed using 8 metrics based on four outcome types: 
{\footnotesize\begin{itemize}
    \item True positive (TP): Delirium patient correctly identified.
    \item False positive (FP): Non-delirium cases misclassified as delirium.
    \item True negative (TN): Non-delirium cases correctly identified.
    \item False negative (FN): Delirium patients misclassified as non-delirium.
\end{itemize}}
Based on these outcomes, we report eight standard evaluation metrics:
{\footnotesize\begin{itemize}
    \item Accuracy: Proportion of correct labels (TP and TN).
    \item Precision (PPV): $\frac{TP}{TP + FP}$, proportion of predicted delirium cases that are correct.
    \item Recall/Sensitivity (TPR): $\frac{TP}{TP + FN}$, proportion of delirium patients correctly identified.
    \item F1 Score: $2 \cdot \frac{\text{precision} \cdot \text{recall}}{\text{precision} + \text{recall}}$, harmonic mean of precision and recall.
    \item Specificity (TNR): $\frac{TN}{TN + FP}$, proportion of non-delirium cases that are correctly identified.
    \item ROC-AUC: Area under the ROC curve, model's ability to distinguish between delirium from non-delirium.
    \item False Alarm (FPR): $\frac{FP}{FP + TN}$, proportion of non-delirium cases misclassified as delirium.
    \item Miss Rate (FNR): $\frac{FN}{FN + TP}$, proportion of delirium cases misclassified as non-delirium.
\end{itemize}}

Rates used to evaluate delirium detection:
{\footnotesize\begin{itemize}
    \item True Delirium Rate = $\frac{\text{Total Number of Positive Admissions}}{\text{Total Number of Admissions}}$.
    \item Estimated Delirium Rate = $\frac{\text{Number of ($TP + FP$)}}{\text{Total Number of Admissions}}$.
    \item Correctly Estimated Delirium Rate = $\frac{\text{Number of $TP$}}{\text{Total Number of Admissions}}$.
\end{itemize}}

\vspace{-0.5pt}
\section{EXPERIMENTS AND RESULTS}
\vspace{-0.5pt}
UC-iML-Delirium models are compared with aML-Delirium and Baseline-Delirium models. Both UC-iML-Delirium and aML-Delirium models utilized multi-source data, while Baseline-Delirium did not. Only the UC-iML-Delirium models incorporated physician expertise.
\vspace{-0.5pt}
\subsection{Experimental Setup}
\vspace{-0.5pt}
The 12 machine learning models, along with hyperparameter tuning and cross-validation, are implemented using the Python package \verb|Scikit-learn| \cite{scikit-learn}. Hyperparameter tuning is conducted using \verb|RandomizedSearchCV| and \verb|GridSearchCV| functions. Cross-validation is employed using \verb|cross_val_score|, \verb|cross_validate| and \verb|cross_val_predict| functions.

The following classifiers are used:
\begin{itemize}
    \item Gradient Boosting: \verb|GradientBoostingClassifier|
    \item AdaBoost: \verb|AdaBoostClassifier|
    \item Neural Network: \verb|MLPClassifier|
    \item Decision Tree: \verb|DecisionTreeClassifier|
    \item k-Nearest Neighbors: \verb|KNeighborsClassifier|
    \item Logistic Regression: \verb|LogisticRegression|
    \item Random Forest: \verb|RandomForestClassifier|
    \item Support Vector Machine: \verb|svm|
    \item Gaussian Naïve Bayes: \verb|GaussianNB|
    \item Linear Discriminant Analysis: \verb|LinearDiscriminantAnalysis|
    \item Quadratic Discriminant Analysis: \verb|QuadraticDiscriminantAnalysis|
    \item Voting Classifier (Soft setting): \verb|VotingClassifier|
\end{itemize}

After data-preprocessing, a decision tree classifier is used to assess the importance of all input features. Recursive feature elimination selected the top 20 features, which are retained for final input.
\vspace{-0.5pt}
\subsection{Experimental Results}
\vspace{-0.5pt}

\begin{table}[ht]
\centering
\caption{\label{lastHDtesting3}Comparison of three types of models: model performance on holdout set 10 (2014.10.01-2015.03.31).} 
\resizebox{0.49\textwidth}{!}{\begin{tabular}{|r|r||r|r|r|}
\hline
\multirow{3}{*}{\textbf{\begin{tabular}[c]{@{}r@{}}Evaluation\\ Metrics\end{tabular}}} & \multirow{3}{*}{\textbf{\begin{tabular}[c]{@{}r@{}}Three\\ Types of\\ Models\end{tabular}}} & \multirow{3}{*}{\textbf{\begin{tabular}[c]{@{}r@{}}Gradient\\ Boosting\\ Classifier\end{tabular}}} & \multirow{3}{*}{\textbf{\begin{tabular}[c]{@{}r@{}}AdaBoost\\ Classifier\end{tabular}}} & \multirow{3}{*}{\textbf{\begin{tabular}[c]{@{}r@{}}Logistic\\ Regression\end{tabular}}} \\
                                                                                       &                                                                                             &                                                                                                    &                                                                                         &                                                                                         \\
                                                                                       &                                                                                             &                                                                                                    &                                                                                         &                                                                                         \\ \hline
\multirow{3}{*}{\textbf{F1 Score}}                                                     & UC-iML-Delirium                                                                             & \textbf{0.719}                                                                                     & 0.704                                                                                   & 0.692                                                                                   \\ \cline{2-5} 
                                                                                       & aML-Delirium                                                                                & 0.652                                                                                              & 0.636                                                                                   & 0.510                                                                                   \\ \cline{2-5} 
                                                                                       & Baseline-Delirium                                                                           & 0.475                                                                                              & 0.467                                                                                   & 0.609                                                                                   \\ \hline
\multirow{3}{*}{\textbf{Accuracy}}                                                     & UC-iML-Delirium                                                                             & \textbf{0.851}                                                                                     & 0.843                                                                                   & 0.841                                                                                   \\ \cline{2-5} 
                                                                                       & aML-Delirium                                                                                & 0.817                                                                                              & 0.811                                                                                   & 0.800                                                                                   \\ \cline{2-5} 
                                                                                       & Baseline-Delirium                                                                           & 0.813                                                                                              & 0.811                                                                                   & 0.804                                                                                   \\ \hline
\multirow{3}{*}{\textbf{Precision}}                                                    & UC-iML-Delirium                                                                             & 0.708                                                                                              & 0.693                                                                                   & 0.700                                                                                   \\ \cline{2-5} 
                                                                                       & aML-Delirium                                                                                & 0.649                                                                                              & 0.641                                                                                   & 0.707                                                                                   \\ \cline{2-5} 
                                                                                       & Baseline-Delirium                                                                           & \textbf{0.896}                                                                                     & 0.894                                                                                   & 0.634                                                                                   \\ \hline
\multirow{3}{*}{\textbf{Recall}}                                                       & UC-iML-Delirium                                                                             & \textbf{0.729}                                                                                     & 0.714                                                                                   & 0.684                                                                                   \\ \cline{2-5} 
                                                                                       & aML-Delirium                                                                                & 0.654                                                                                              & 0.632                                                                                   & 0.398                                                                                   \\ \cline{2-5} 
                                                                                       & Baseline-Delirium                                                                           & 0.323                                                                                              & 0.316                                                                                   & 0.586                                                                                   \\ \hline
\multirow{3}{*}{\textbf{Specificity}}                                                  & UC-iML-Delirium                                                                             & 0.894                                                                                              & 0.888                                                                                   & 0.896                                                                                   \\ \cline{2-5} 
                                                                                       & aML-Delirium                                                                                & 0.875                                                                                              & 0.875                                                                                   & 0.941                                                                                   \\ \cline{2-5} 
                                                                                       & Baseline-Delirium                                                                           & \textbf{0.987}                                                                                     & \textbf{0.987}                                                                          & 0.880                                                                                   \\ \hline
\multirow{3}{*}{\textbf{ROC-AUC}}                                                      & UC-iML-Delirium                                                                             & 0.902                                                                                              & \textbf{0.905}                                                                          & 0.896                                                                                   \\ \cline{2-5} 
                                                                                       & aML-Delirium                                                                                & 0.882                                                                                              & 0.875                                                                                   & 0.831                                                                                   \\ \cline{2-5} 
                                                                                       & Baseline-Delirium                                                                           & 0.834                                                                                              & 0.825                                                                                   & 0.820                                                                                   \\ \hline
\multirow{3}{*}{\textbf{False Alarm}}                                                  & UC-iML-Delirium                                                                             & 0.106                                                                                              & 0.112                                                                                   & 0.104                                                                                   \\ \cline{2-5} 
                                                                                       & aML-Delirium                                                                                & 0.125                                                                                              & 0.125                                                                                   & 0.059                                                                                   \\ \cline{2-5} 
                                                                                       & Baseline-Delirium                                                                           & \textbf{0.013}                                                                                     & \textbf{0.013}                                                                          & 0.120                                                                                   \\ \hline
\multirow{3}{*}{\textbf{Miss Rate}}                                                    & UC-iML-Delirium                                                                             & \textbf{0.271}                                                                                     & 0.286                                                                                   & 0.316                                                                                   \\ \cline{2-5} 
                                                                                       & aML-Delirium                                                                                & 0.346                                                                                              & 0.368                                                                                   & 0.602                                                                                   \\ \cline{2-5} 
                                                                                       & Baseline-Delirium                                                                           & 0.677                                                                                              & 0.684                                                                                   & 0.414                                                                                   \\ \hline
\end{tabular}}

\end{table}

Models are trained on the first 9 six-month time segments (April 2010--September 2014) using hyperparameter tuning with 5-fold cross-validation. We evaluate 12 classifiers under the same protocol and report the top-performing models in the main text; remaining model-level results are omitted due to space constraints. The prospective holdout test set is the final temporal segment (October 2014--March 2015, $n=104$), used to simulate future prediction (Fig.~\ref{tt2}). The dataset contained 3,862 labeled admissions across 10 sequential temporal segments, with delirium prevalence ranging from 23\% to 28\%.

The results can be viewed in TABLE~\ref{lastHDtesting3}. The UC-iML-Delirium achieved the best overall discrimination (accuracy, recall, ROC-AUC, and F1-score), while Baseline-Delirium showed higher precision and specificity. Beyond the holdout summary in TABLE~\ref{lastHDtesting3}, we further evaluate temporal stability using rolling validation windows and assessed subgroup consistency by sex. 

Temporal stability analysis of the three top algorithms indicated that UC-iML-Delirium is more consistent across all six metrics. For GBC, UC-iML-Delirium achieved an F1 score of $0.701\pm0.024$ (range: 0.646--0.719) across rolling windows, compared with $0.749\pm0.136$ for aML-Delirium and $0.402\pm0.046$ for Baseline-Delirium, demonstrating robust performance over the five-year period. Model performance is statistically comparable between male and female patients, with both groups’ F1 scores within the 95\% confidence interval [0.680, 0.758].

\begin{figure}[H]
    \centering
    \subfloat[SHAP diagram of feature importance for global model.\label{fig:s1}]{
        \includegraphics[width=0.80\linewidth]{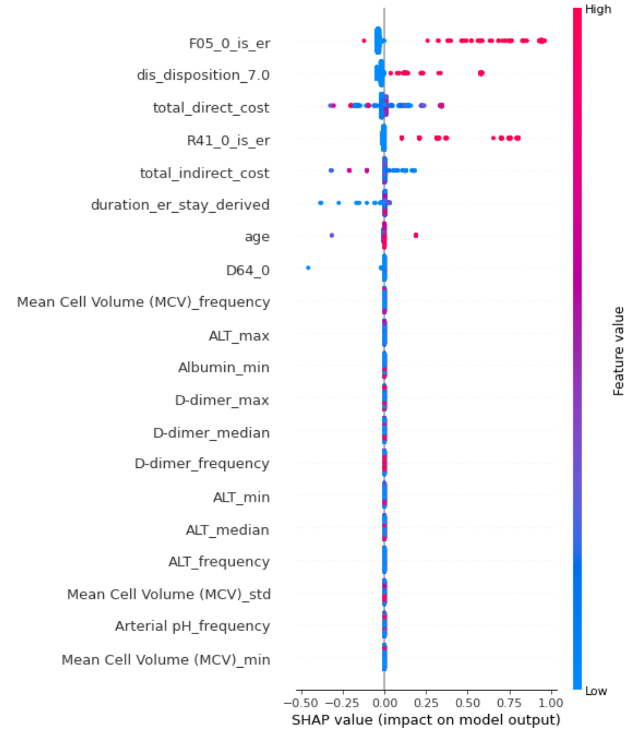}
    }
    \hfill
    \subfloat[Top 10 selected risk factors of delirium.\label{fig:risk_factors}]{
        \begin{minipage}[b]{0.80\linewidth}
            \centering
            \begin{tabular}{@{}p{5.0cm}cc@{}}
                \toprule
                \textbf{Risk Factor} & \textbf{Weight} & \textbf{Sign} \\
                \midrule
                text-derived delirium probability & 8.846 & + \\
                Diagnosis code F05 (Emergency Room) & 1.991 & + \\
                Number of antipsychotics orders & 0.707 & + \\
                Admit via ambulance & 0.531 & + \\
                Discharge to position & $-0.451$ & $-$ \\
                Lactate Arterial (minimum) & 0.387 & + \\
                Diagnosis code F05 & 0.365 & + \\
                INR (minimum) & $-0.305$ & $-$ \\
                Diagnosis code K92 (ER) & 0.290 & + \\
                Diagnosis code I50 (ER) & 0.274 & + \\
                \bottomrule
            \end{tabular}
        \end{minipage}
    }
    \caption{Model interpretation results: (a) Feature importance visualization using SHAP values, and (b) Quantitative risk factors with corresponding weights and directional effects.}
    \label{fig:combined_analysis}
\end{figure}

\begin{table}[!t]
\centering
\caption{Model performance on independent validation datasets. Phase 1: prospective holdout test on last time segment (October 2014--March 2015, n=104). Phase 2: temporal validation on data collected 3--5 years after training period (2018--2020, n=358). All models trained on Phase 1 data (n=3,353). GBC = Gradient Boosting Classifier; RF = Random Forest; LR = Logistic Regression; ABC = AdaBoost Classifier.}
\label{tab:independent_validation}
\setlength{\tabcolsep}{10pt}
\small
\begin{tabular}{l|cccc}
\toprule
\rowcolor{gray!30}
\multicolumn{5}{c}{\textbf{Phase 1: Prospective Holdout Test (n=104)}} \\
\midrule
\rowcolor{gray!20}
\textbf{Metric} & \textbf{GBC} & \textbf{RF} & \textbf{LR} & \textbf{ABC} \\
\midrule
\rowcolor{blue!5}
Accuracy          & 0.779 & 0.702 & 0.731 & 0.760 \\
Precision         & 0.778 & 0.652 & 0.692 & 0.741 \\
\rowcolor{blue!5}
Recall            & 0.553 & 0.395 & 0.474 & 0.526 \\
F1 Score          & 0.646 & 0.492 & 0.562 & 0.615 \\
\rowcolor{blue!5}
ROC-AUC           & 0.866 & 0.805 & 0.809 & 0.827 \\
Specificity       & 0.909 & 0.879 & 0.879 & 0.894 \\
\rowcolor{blue!5}
False Alarm       & 0.091 & 0.121 & 0.121 & 0.106 \\
Miss Rate         & 0.447 & 0.605 & 0.526 & 0.474 \\
\midrule
\rowcolor{gray!30}
\multicolumn{5}{c}{\textbf{Phase 2: Temporal Validation (n=358)}} \\
\midrule
\rowcolor{gray!20}
\textbf{Metric} & \textbf{GBC} & \textbf{RF} & \textbf{LR} & \textbf{ABC} \\
\midrule
\rowcolor{blue!5}
Accuracy          & 0.827 & 0.827 & 0.782 & 0.813 \\
Precision         & 0.875 & 0.929 & 0.677 & 0.877 \\
\rowcolor{blue!5}
Recall            & 0.574 & 0.533 & 0.689 & 0.525 \\
F1 Score          & 0.693 & 0.677 & 0.683 & 0.656 \\
\rowcolor{blue!5}
ROC-AUC           & 0.900 & 0.883 & 0.874 & 0.885 \\
Specificity       & 0.958 & 0.979 & 0.831 & 0.962 \\
\rowcolor{blue!5}
False Alarm       & 0.042 & 0.021 & 0.169 & 0.038 \\
Miss Rate         & 0.426 & 0.467 & 0.311 & 0.475 \\
\bottomrule
\end{tabular}
\end{table}

\vspace{-0.5pt}
\subsection{SHAP Experimental Results}
\vspace{-0.5pt}


Model performance differed between Phase~1 and Phase~2 (TABLE~\ref{tab:independent_validation}). Fig.~\ref{fig:s1} shows that both engineered variables and diagnosis-related signals contributed to prediction, with \verb|F05_0_is_er| among the strongest features. This pattern supports clinical face validity, but it also suggests that the current model is better interpreted as delirium detection support within routine hospital data than as a pure early-warning model based only on pre-diagnostic signals. These diagnoses are independent of the final delirium label assigned via administrative chart review \cite{saczynski2014tale}.

\begin{figure}[t]
    \centering
    \includegraphics[width=0.95\columnwidth]{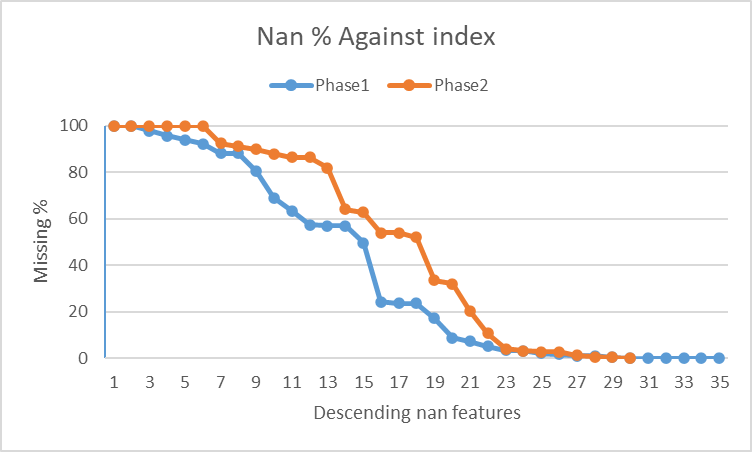}
    \caption{Distribution of missing percentages in the original features for phases 1 and 2.}
    \label{fig:22}
\end{figure}

Fig.~\ref{fig:22} summarizes feature-level missingness in Phases~1 and~2. Phase~2 exhibits higher missingness, suggesting that performance differences are unlikely to be driven by improved data completeness. TABLE~\ref{tab:independent_validation} also shows that the model under-estimated delirium prevalence in both phases and achieved only modest recall. Together with the lower delirium prevalence in Phase~2, these results are consistent with increased class imbalance and motivate calibration-aware thresholding for screening-oriented use.

    

\vspace{-0.5pt}
\section{IDENTIFYING DELIRIUM RISK FACTORS: ENGINEERED FEATURES DOMINATE}
\vspace{-0.5pt}
Risk factor analysis helps optimize prevention and treatment by identifying key variables \cite{pett2003making, aussem2012analysis}. 
This study applied ML methods to integrate multi-source data, using feature engineering to create variables such as natural language processing (NLP)-derived features from radiology reports and flag variables for dimensionality reduction in lab tests. Logistic regression analyzes the relationship between these features and delirium risk \cite{zaal2015systematic, miravitlles2000factors}. 

The logistic regression model is implemented with \\ hyperparameter tuning and 5-fold cross-validation in \verb|Scikit-learn| \cite{scikit-learn}, using \verb|RandomizedSearchCV| and \verb|GridSearchCV| for tuning, and \verb|cross_val_score|, \verb|cross_validate|, and \verb|cross_val_predict| for validation. The logistic regression classifier is conducted using the \verb|theLogisticRegression| function.

\vspace{-0.5pt}
\subsection{Experimental Results}
\vspace{-0.5pt}
The model is trained on the first 9 time segments, with the last 6 months of GEMINI data used as the testing set. Accuracy and F1 score are evaluated. The training accuracy is 0.874 $\pm$ 0.033, with a corresponding F1 score of 0.841, while the testing accuracy is 0.729 $\pm$ 0.084, with a corresponding F1 score of 0.692.



Fig.~\ref{fig:risk_factors} summarizes the top delirium risk factors identified by logistic regression, including each feature's coefficient and directionality. The suffix \verb|_ER| denotes emergency-room variables, and \verb|_min| denotes the minimum value of a laboratory test during the hospital stay.

\vspace{-0.5pt}
\subsection{Summary}
\vspace{-0.5pt}
The top 10 delirium risk factors included both derived and original features. Derived features are from radiology reports, antipsychotic medication orders, lactate arterial lab values, and INR during the hospital stay. Original features included key diagnosis codes (F05, K92, I50) and patient discharge disposition at discharge.

Understanding delirium risk factors is crucial for improving management. Feature engineering and data integration revealed new risk factors, with the UC-iML-Delirium model showing significant results for delirium management in practice. Future research will focus on refining delirium detection and treatment.

\vspace{-0.5pt}
\section{DISCUSSION}
\vspace{-0.5pt}
The UC-iML-Delirium method generally outperformed other models across evaluation measures, excelling true positives such as recall and F1 score. While its performance is weaker on precision, the high variability in precision across time suggests lower reliability in this context. It balanced true positive detection and false positive avoidance, with improvements seen in phase 2 due to dataset imbalance, where 75 \% of admissions are non-delirium. To address the precision-recall trade-off, the F1 score, which balances both, is used. The UC-iML-Delirium method outperformed other schemes across all time segments when evaluated using the F1 score.
The integration of physician input into iML-Delirium enhances model performance and usefulness, with potential for broader application in medical decision-making.

A limitation of this study is using a common set of features across phases 1 and 2 to assess model stability. While better performance has been reported in other studies using the full dataset, this study focused on evaluating stability and drift, rather than optimization.

\vspace{-0.5pt}
\section{CONCLUSION}
\vspace{-0.5pt}
Delirium is a prevalent, preventable, and treatable neurocognitive disorder associated with poor outcomes when untreated. Its acute onset and fluctuating symptoms make it difficult to detect, underscoring the need for automated delirium identification to improve care quality.
This research demonstrates the stability of ML model performance using data from six Toronto hospitals, effectively predicting delirium presence even when tested on data from one hospital years later. Model performance slightly improved in phase 2, likely due to a reduction in the positive delirium rate, rather than improved data quality, as phase 2 contained more missing values.
Incorporating physician expertise within the UC-iML-Delirium  approach through human-in-the-loop design significantly improved prediction accuracy compared to standard ML methods. Temporal stability analysis revealed that the UC-iML-Delirium  model outperformed others, highlighting the benefits of combining physician expertise with machine learning throughout the development process. Neither feature engineering nor physician expertise alone achieved comparable results.
These findings emphasize the value of UC-iML-Delirium  with human-in-the-loop design in predicting critical healthcare outcomes, demonstrating that continuous physician involvement enhances both model performance and clinical trust. Future research should focus on: a) better integrating physician expertise into ML predictions, and b) refining the UC-iML-Delirium methodology for improved clinical applicability and integration.

\vspace{-0.5pt}
\section*{Acknowledgment}
\vspace{-0.5pt}
The authors would like to thank Dr. Amol Verma and Dr. Fahad Razak for providing access to the GEMINI dataset and for their invaluable clinical expertise and guidance, which are essential to the development and validation of this work.
\bibliographystyle{plain}
\bibliography{backup}









\end{document}